\pdfoutput=1
\documentclass[lettersize,journal]{IEEEtran}
\usepackage{amsmath,amsfonts}
\usepackage{array}
\usepackage[caption=false,font=normalsize,labelfont=sf,textfont=sf]{subfig}
\usepackage{textcomp}
\usepackage{stfloats}
\usepackage{url}
\usepackage{verbatim}
\usepackage{graphicx}
\usepackage{cite}

\usepackage{arydshln}
\usepackage[normalem]{ulem}
\usepackage{algorithm}
\usepackage{algpseudocode}
\usepackage{float}
\usepackage[shortlabels]{enumitem}
\usepackage{amssymb}
\usepackage{amsthm}
\theoremstyle{definition}
\newtheorem{definition}{Definition}[section]
\newtheorem{statement}{Modeling statement}[section]

\algdef{SE}[DOWHILE]{Do}{doWhile}{\algorithmicdo}[1]{\algorithmicwhile\ #1}%

\DeclareMathOperator*{\argmin}{arg\,min}

\begin{document}

\title{Potential gains of communication-compute-control co-design 
based performance optimization methods in cyber-physical systems}

\author{S\'andor R\'acz and Norbert Reider, Ericsson Research, Hungary

\thanks{E-mail: sandor.racz@ericsson.com, norbert.reider@ericsson.com}
}



\maketitle

\begin{abstract}

In this paper we propose and quantitatively evaluate three performance optimization methods that exploit the concept of communication-compute-control co-design by introducing awareness of communication and compute characteristics into the application logic in different ways to improve overall system performance. 
We have implemented a closed-loop control of a robotic arm over a wireless network where the controller is deployed into an edge cloud environment. When implementing an industrial system that leverages network and cloud technologies, the level of determinism of the control application can be decreased by nature. 
This means that some imperfections may be introduced into the control system, and the closed-loop control in substance changes to open-loop during disturbances. 
We aim to improve the performance of these open-loop control periods by applying methods that can compensate for the imperfections statistically or in a guaranteed way.

We demonstrate that co-design-based application improvements with minimal dependencies on the underlying technologies can already yield an order of magnitude gain when it comes to the accurate execution of the robot trajectories during the open-loop control periods.
Furthermore, by combining the proposed methods, the performance improvements add up and can produce up to 45\% shorter trajectory executions compared to individual evaluations.

\end{abstract}


\section{Introduction}

\IEEEPARstart{T}{he} next generation of Cyber-Physical Systems (CPS) are about to transform industries by taking advantage of innovative technologies such as wireless networks, cloud computing, digital twins, advanced control solutions and Artificial Intelligence (AI). These technologies are driving the evolution of CPS into highly intelligent systems that impact various sectors, improve overall functionality and efficiency, and enhance the level of automation \cite{Song2016}. Modern communication networks provide reliable and fast communication between control components and foster the penetration of Networked Control Systems (NCS). NCS connect control components within a control loop over a network and have been applied in many areas, such as space environments exploration, industrial automation, robots, aircraft, automobiles, manufacturing plant monitoring, remote diagnostics and troubleshooting, and teleoperations \cite{Zhang2020}. In Wireless Networked Control Systems (WNCS), sensors and actuators are connected through wireless connections to their controllers. WNCS are strongly related to CPS since these emerging techniques deal with real-time control of physical systems over the wireless networks \cite{Park2018}. 

Introducing wireless network simply as a replacement of cables poses huge challenges on the underlying wireless communication systems \cite{Wang2024}. 
Most of the conventional control systems of today deal with wired communication medium with the assumption of close to zero delay, jitter and packet loss, as well as ultra-high data rates and reliability \cite{Park2018}, \cite{Zhao2019}. 
Some wireline industrial communication protocols can be configured to tolerate a few numbers of communication cycles without valid data reception before a communication issue is reported in order to accommodate short-term failures, typically up to the range of 10~ms.

There have been many efforts to design wireless technologies such that they approach the performance of wired communication systems in terms of data rate, latency, reliability and resilience using different techniques such as massive Multiple-Input Multiple-Output transmission, advanced channel prediction, ultra-dense small cells, and so on. Such solutions typically require extensive network resources or they are simply costly, and in some cases, it is still infeasible to fulfill the very strict requirements of conventional control applications designed to work over cables \cite{Willig2008}, \cite{Gungor2009}, since the dynamic nature of wireless channels can contradict the determinism implemented in legacy control systems \cite{Zhao2019}. That is, perfect radio channel cannot be inherently guaranteed. 

Many industrial applications typically enter emergency state if communication problems are detected and the whole system is stopped to prevent damage and ensure safety, which impacts not only one application but also the entire production \cite{Severson2016}, \cite{Reis2017}. Therefore, simple cable replacement is not a straightforward way forward for many conventional applications using wireless communication, in particular not for critical applications or where resource efficiency is of importance. 

However, there are still several benefits why wireless communication should be applied in control systems. The appearance of WNCS is a key enabler for improved flexibility, mobility and robustness in CPS and brings in great innovation potential. 
Furthermore, wireless networks are significantly simpler in their deployments or to retrofit connectivity onto deployed systems.
Many control systems involve motion or mobility where wired connectivity is impractical \cite{Noor2023}.
From application perspective, WNCS are in high demand in several areas such as industrial automation \cite{Willig2008}, automotive \cite{Sadi2013}, avionics \cite{Itur2013} and building management \cite{Witrant2010}.

Moreover, there is a growing interest to offload functionality from physical systems to cloud environments where scalable access to enormous compute capabilities is enabled and opens for data-driven and AI-based optimizations \cite{Yu2018}. The computation power of the existing purpose-specific hardware can be insufficient to process the large amount of data, execute the modern control strategies such as Model Predictive Control (MPC), and train and inference the AI models in the low-level control loops, in real-time and at scale \cite{LeeJay2020}, \cite{Peres2020}. Offloading to cloud also facilitates the realization of practical collaborative use cases where centralized information is required \cite{Knepper2013}.

Cloud-native applications offer the prospect of greater flexibility, reuse, availability, and reliability with lower latencies \cite{Skarin2018}.
Introducing cloud execution for monolith applications is considered to be inefficient and prevents realizing the advantages of cloud execution, instead the move is towards cloud-native applications.
Many of today’s control systems are still deployed as monolithic implementations on dedicated or tuned hardware, typically applying real-time operating systems. Such a setup is non-modular, less extensible and limits the ability to self-adapt \cite{Skarin2018}. As an example, today’s low-level functionalities such as closed-loop control, are still executed in purpose-specific hardware such as Programmable Logic Controllers (PLCs), industrial PCs, and robot controllers and they are connected by wired networks \cite{Sehr2021}.
Consequently, as a first step, decomposing the control program into smaller functions for reliable and scalable cloud execution is needed to develop fully cloud-native control applications \cite{Gil2021}. 

The introduction of wireless networking and cloud computing into CPS as a combination offers benefits that can improve flexibility, resilience and scalability in order to drive the evolution in many verticals such as smart manufacturing, automotive industry or agriculture. Several EU projects focus on to demonstrate the capabilities of 5G/6G networks combined with cloud computing for different verticals. For instance, the 5G for Smart Manufacturing project \cite{5GSMART} has already been concluded on the validation of 5G network in factory automation. The Deterministic6G project \cite{Det6G} is still ongoing to develop the next generation of deterministic networking capabilities within the context of 6G networks.

On the other hand, applying advanced technologies in control systems such as wireless and cloud computing can decrease the deterministic behavior of the system.
Cloud environment can introduce some imperfections from determinism perspective, e.g., in keeping delay deadlines or handling infrastructure failures \cite{Hu2012}. As discussed above, the complete elimination of the sources of such imperfections can not only be technically challenging but also less motivated from resource efficiency, sustainability and cost perspectives. 
Thus, some residual imperfection can remain in the system by design \cite{Chien2019}, \cite{Janković2022}.
Moreover, control application logic itself may abandon deterministic behavior intentionally and temporarily to handle practical cases such as control overload, bandwidth usage or energy efficiency, and thus deviates from its strictness to ensure overall system efficiency, responsiveness, and stability \cite{Yan2021}.

\subsection{5G technology}
The 3GPP 5G technology \cite{3GPP5G} is the first wireless telecommunication standard that is explicitly targeting Industrial Internet of Things via dedicated features such as the Ultra-Reliable Low Latency Communication and the massive Machine Type Communication for large scale connectivity of devices. The 5G also supports the deployment of dedicated private networks and provides interfaces and solutions to interconnect with the wireline standards of deterministic packet networking protocols such as the IEEE 802.1 Time Sensitive Networking \cite{TSN} and IETF Deterministic Networking \cite{DetNet}. Moreover, 5G enables network programmability through application programming interface (API) exposure to configure the network and retrieve a variety of network information. Other features of 5G such as traffic prioritization via the wide ranges of QoS settings, customizable transmission slots to adapt to uplink heavy applications, the dynamic spectrum sharing, as well as the support of Reduced Capability devices can facilitate its adoption into the industrial domain \cite{5Gbook}.
The 5G technology provides a standardized and complete ecosystem and thus, it can be a good candidate to be selected by industrial players as a future-proof technology for use cases where flexible, high performing and reliable connection is required.
%
%

\subsection{Related work}

Co-designing communication, compute and control technologies have lately garnered increased interest, though it has previously been explored from various perspectives such as the cross-layer optimization methods. In this section, we highlight our additional contributions while we explore some relevant prior work for different technology areas.

\subsubsection{Related work from the area of NCS and WNCS}

The authors of \cite{Zhang2020} present a survey of trends and techniques in networked control systems from the perspective of \emph{‘control over networks’}. They concluded that a co-design scheme is expected for an NCS by simultaneously taking \emph{‘control of networks’} and \emph{‘control over networks’} into account which enables suitable communication protocol design, ensures good quality of service (QoS) of communication networks, as well as notable control strategies can be provided such that the NCS achieve desirable control performance to carry out expected control tasks.

In \cite{Park2018}, the authors perform an exhaustive literature review of WNCS design and optimization solutions where the main challenge is to jointly design the communication and control systems considering their tight interaction to improve the control performance and the network lifetime. Moreover, the critical interactive variables of communication and control systems, including sampling period, message delay, message drop, and energy consumption are also discussed. 

The authors of \cite{Wang2024} present a comprehensive survey of WNCS from the communication perspective. They discuss appropriate WNCS architecture and other topics, such as sensing strategy design under energy and bandwidth constraints, state estimation problems in the presence of imperfect channels, and control approaches for WNCS performance.

Several papers such as \cite{Wildhagen2022} investigate the effect of maximum delay and propose solutions to compensate that. Advanced control approaches such as MPC have also been extended to handle lossy network \cite{Umsonst2024} or imperfect network connections for safety-critical teleoperation \cite{UR5meas}.

The solutions mentioned above for NCS and WNCS are usually handle the problem of co-design as joint optimization problems, or rely on low-level, control theory-based approaches with the focus on stochastic variations. The control theory-based solutions can compensate for delays, small jitters and shorter outages effectively. However, decomposing the problem of handling network imperfection into multiple levels of mechanisms that are on top of each other can be beneficial. Thus, we handle outages directly case-by-case, and the packet-level stochastic variations (e.g., per packet delay, jitter and random packet loss) are assumed to be compensated by state-of-the art control theory-based solutions. Our contribution focuses on those imperfections that are not compensated by prior art methods (typically larger outages) and thereby complementing the existing solutions.

\subsubsection{Related work from cloud domain}

The concept of Cloud-Fog Automation has been introduced in \cite{Lee2020} and \cite{Jin2024} as a new design paradigm for factory automation systems to move progressively from a classical hierarchical pyramid architecture toward a more cloud-based automation architecture. \cite{Lv2024} discusses how cloud/fog-based virtualization and converged communication networks can allow for flexible deployment of automation use cases and evaluate the impacts introduced by the less deterministic infrastructure from control perspective.

In modern industrial applications, the migration of higher-layer control to the cloud has already become a standard practice \cite{Jin2024}. In conjunction with this, cloud robotics has recently emerged as a vibrant area of research, combining cloud computing and robotics through advancements in cloud technologies and wireless networks \cite{Saha2018}.

Several works deal with the challenges and opportunities of cloud-based control of robots as discussed in \cite{Kehoe2015} where the authors group the papers around four potential benefits of the cloud: Big data, Cloud computing, Collective robot learning and Human computation. 
The authors highlight that new algorithms and methods are needed to cope with time-varying network latency and QoS. Faster data connections, both wired Internet connections and wireless standards are reducing latency, but algorithms must be designed to degrade gracefully when cloud resources are very slow, noisy, or unavailable.

The containerization as an emerging, efficient and truly cloud-native alternative to traditional virtualization is gaining widespread popularity \cite{Liu2021} since it has shown clear advantages over traditional Virtual Machines (VMs) \cite{Bentaleb2022}. The paradigm of device-edge-cloud continuum \cite{Savaglio2023} leverages a distributed cloud-edge infrastructure to execute applications on different hardware platforms and locations based on where they achieve better performance.

Since the aforementioned initiatives tries to optimize the cloud deployment of the applications in various ways, they utilize the co-design methodology of the application and the compute domains in some form. We also apply containerization techniques throughout the paper as the cloud-native execution platform of our robotic application.

\subsubsection{Pragmatic co-design approaches}

The authors of \cite{Qiao2023} show that the communication–control co-design can lower the requirements on the coding rate and lower the consumption of wireless resources in a cloud control Automated Guided Vehicles (AGVs) example. The proposed co-design method also achieves better results in the probability of system instability and the number of admissible AGVs. 

Another case study is presented in \cite{Lyu2023} to demonstrate the advantages and differences of co-designed system compared to the conventional independent design. Therefore, a latency-aware wireless control framework is presented which investigates the impacts of the commonly used industrial wireless networks on the control performance parameters using a ball-and-beam time-critical balancing control system and provides more restricted position error results compared to the traditional cascaded PID controller, improving the robustness of the system.

The work presented in \cite{Lv2023} implements a network hardware-in-the-loop simulation framework to effectively and efficiently investigate the impacts of wireless on robot control under real network conditions and then to improve the design by employing correlation analysis between communication and control performances.

The authors of \cite{Liu2022} and the references therein applied cross-layer design approaches for various use cases to enhance overall system performance. The solution in \cite{Liu2022} applied in an Unmanned Aerial Vehicle (UAV) use case to address the unstable transmission and heavy load issues of wireless networks. The proposed framework showed significant performance gain of the whole UAV system achieved by the cross-layer design.

Although these studies provide significant value by adopting a realistic use case, they primarily examine specific co-design algorithms that typically center around control theory-based solutions or applied between different network layers. We aim to complement these solutions by proposing mechanisms on higher level by the means of advanced application-level algorithms that exploits insights from other technologies (e.g., from the communication or cloud domains).

\subsubsection{Conceptual and theoretical work on co-design}

The co-design of communication and control applications are studied in \cite{Zhao2019}, where the authors discuss fundamental design capabilities needed to realize real-time control in future wireless networks, with primary emphasis given to communication and control since they tightly interact with each other.

A decent summary of existing literature on co-design is provided in \cite{Qiao2023}. The authors highlight that majority of the work formulate joint optimization problems to address the co-design of communication and control systems. The existing co-design methods are primarily categorized into two types of problems: control optimization under communication constraints and communication optimization under control constraints. Certain parameters simultaneously impact both systems, forming the core of communication–control co-design.

The solutions outlined above approach the co-design problem analytically, aiming to achieve complex, yet optimal solutions. In this paper we consider a practical and common use case instead, namely the cloud control of a robotic arm, to propose and quantitatively evaluate some co-design-based performance optimization solutions. We note that the proposed solutions can be applied not only for robotic arms, but for any system where motion planning is applied.

\subsection{Relevance of co-design}

The determinism of control systems is to a large extent an implementation choice.
It is influenced by the architecture of the system, for instance, the design of the control algorithms, the choice of the communication and compute solutions. By changing the way control systems are implemented, the level of determinism may be different/relaxed. Moreover, the requirements may also be varying depending on the actual scenario, over time, etc. This flexibility opens up new opportunities where co-design of communication and application is expected to play a key role \cite{Zhao2019}.
The co-design also facilitates to mitigate the effect of the residual imperfection of system components.

The tight coupling among communication, compute and control requires to treat these technologies as one system to design a fully integrated system that achieves improved performance while consuming adequate wireless and compute resources \cite{Zhao2019}. 
The emerging relevance of co-design is also promoted by trends in CPS, namely the introduction of wireless networking for flexible deployments, as well as the adoption of cloud computing for reliable and scalable computing.
Rewriting control applications to consider wireless communication aspects is a significant demand on industrial partners. However, re-implementing applications is also a need from cloud computing point of view.

\subsection{Scope and contributions}


We assume that the robotic application, including also the closed-loop control part, is offloaded to an edge cloud infrastructure \cite{Qui2020} and wireless network is applied as the communication medium between the application and the robot. 
We present and quantitatively evaluate our proposed methods in a use case where a UR5e industrial robotic arm \cite{UR5e} is controlled from edge cloud using a closed-loop velocity controller over a 3GPP Private 5G Standalone Release-16 network \cite{5GSA}. Before we culminated in this testbed, we conducted explorative measurements as detailed in \ref{appendix:background}.
However, the proposed methods can be adapted for any system where trajectories are used for describing motion and they are not dependent on specific robots, networks or cloud platforms. For instance, they can be applied for mobile robots as well when control is over wireless and accurate motion is required.

We aim to add improvements in algorithms not by other means, such as by adding extra network resources for densification or by duplicating the control stack or by sending complete trajectories over the network, since those solutions are usually resource intensive and the network resources are relatively expensive \cite{Chien2019}. 

We propose methods to improve application-level performance. These methods exploit the freedom of adapting the underlying mechanisms as part of the communication-compute-control co-design concept.
The intention of the methods is to reduce the deviation from the planned trajectory by mitigating the effect of the residual imperfection of the underlying components.

\section{System model}

\subsection{Assumptions}
\label{sec:Assumptions}

We introduce the interpretation of \emph{imperfections} in this context and define our statement on the \emph{modeling of these imperfections}, since they serve as the basis of the proposed methods.

\begin{definition}[Imperfections]
By imperfections we mean any intentional or unexpected operation from the application perspective which decreases the deterministic behavior of the application logic. Imperfections can be caused by several sources such as
\begin{itemize}
    \item the nondeterministic behavior of the underlying wireless connection;
    \item the possible glitches in the cloud execution environment (e.g., process migration), 
    \item stochastic behavior of e.g., AI-based services such as computer vision, and
    \item the effect of other deliberate, higher application-level goals such as improved resource or energy efficiency.
\end{itemize}
\end{definition}

\begin{statement}[Imperfections are modelled as ``gaps'']
We model the \emph{imperfections} as the bursts of missing or late control commands that we refer to as \emph{gaps} throughout the paper. 

Such conservative modeling enables us to
\begin{itemize}
    \item evaluate the performance aspects of the worst-case scenarios;
    \item develop more compact methods with less direct dependencies on the underlying technologies;
    \item configure the methods easily, and
    \item cover all kind of imperfections in the same, simple way making the complexity tractable.
\end{itemize}
\end{statement}

By using this modeling approach, we follow the Occam's razor principle which recommends searching for explanations constructed with the smallest possible set of elements. It is also known as the law of parsimony.

In the proposed methods, we assume that the maximum length of a gap is known and used as input for the particular method. Moreover, we presume that gaps occur relatively seldom, i.e., they are not consecutive, and the controller can recover from the previous one and the transient period faded out before the next one happens. This assumption is typically met in case of modern technologies such as the 5G network and an edge cloud environment tuned for industry use cases \cite{Törngren2021}.

\begin{figure*}[!t]
\centering
    \centering
    \includegraphics[width=1.0\textwidth]{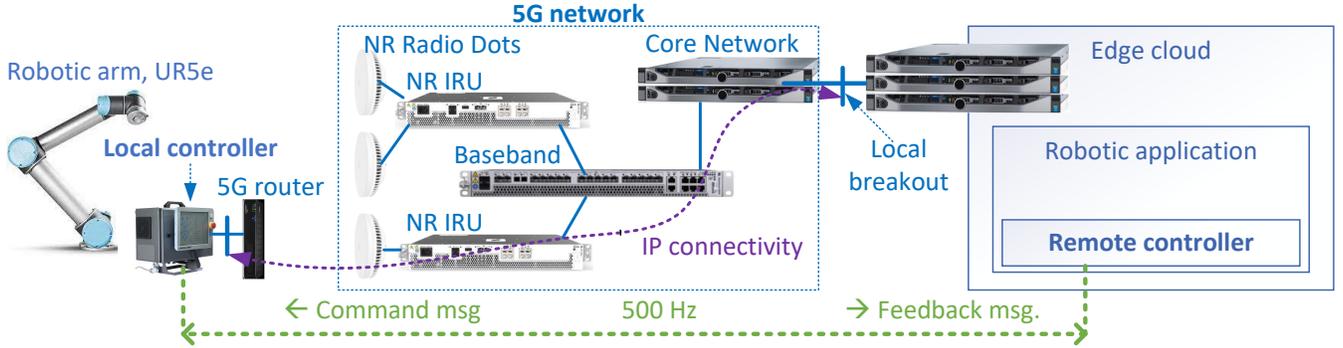}
    \caption{Illustration of the functional system architecture combined with the 5G SA private network deployment used for the measurements}
    \label{fig:architecture}
\hfill
\end{figure*}

We assume intelligent underlying components that have the capability of self-observability, e.g., network can measure delay, jitter, loss and outage gaps for a long period.
However, this information could also come from the network exposure service of the 5G which is expected to provide such information, e.g., using predictions and offered via exposure APIs \cite{5GACIAExposure}. Network exposure is part of the 5G programmability feature which opens up some network capabilities and services towards third-party developers to build network-aware applications. For instance, the application can request a specific Quality of Service for a connection, requesting notification of QoS changes, as well as monitoring the QoS of a connection \cite{5GACIAQoS}. Thus, 5G enables \emph{dynamic information exchange} between the network and the application. These capabilities are the first but essential steps of the journey towards the fully programmable 6G networks \cite{6GEricsson}.

\subsection{System architecture}

Figure \ref{fig:architecture} illustrates our functional system architecture where a UR5e robotic arm is controlled from an edge cloud environment over a 5G network. 

The 5G network operates in locally licensed mid-band spectrum at $3.64–3.66$ GHz (5G band n$78$) and provides enhanced Mobile Broadband service. It is a similar setup that is used in \cite{Ansari2022} at the ``Reutlingen Trial Site'' with the only difference of using a bandwidth of $20$ MHz instead of $100$ MHz and a slightly different spectrum. 
The 5G Radio Access Network is realized with Ericsson Radio Dot System consisting of three main components: Baseband Unit, New Radio (NR) Indoor Radio Unit (IRU) and NR Radio Dots \cite{RDS}. The edge cloud servers are connected to the Core Network via Local breakout connection. This configuration enables us to route the data traffic directly to a local network bypassing the central core network components to reduce latency.

Our purpose-build robotic application runs in an edge cloud container. Its main functionalities are trajectory generation from waypoints, loading serialized trajectories and trajectory executions using a closed-loop feed-forward joint velocity controller. In addition to this, it implements the proposed methods and conducts detailed measurements.

The local controller is the legacy controller box of UR5e robot (also used in \cite{UR5meas}), where incoming speed commands are translated into actual current values to be used for the respective servo motors. This controller is also responsible for providing the feedback messages including the description of the internal state of the robot (actual positions, speeds, temperature, etc.).

Appendix \ref{appendix:LTE_RTT} contains some measurements on delay characteristics of the command and feedback messages between the local and the remote controllers.

\section{Proposed performance optimization methods}
\label{sec:proposedMethods}

\begin{figure}[ht]
    \centering
    \includegraphics[scale=0.46]{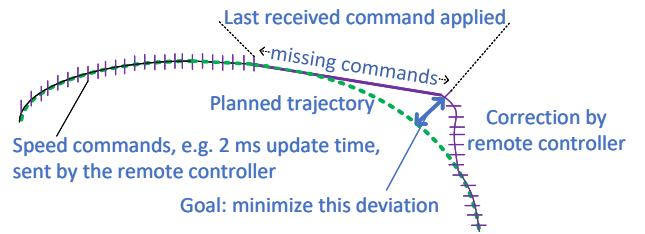}
    \caption{Illustration on the impact of a gap on the trajectory execution}
    \label{fig:pidIllustration}
\end{figure}

Our methods aim to decrease the deviation from planned trajectory during a gap as Figure \ref{fig:pidIllustration} illustrates. 

During a gap, the local controller supervises the robot movement relying on the latest command received from the remote controller. This is a built-in behavior of the local controller of a typical industrial robotic arm such as the UR5e robot \cite{URScript}. 
This open-loop control period can cause transient deviations from the planned trajectory. After the gap is over, the closed-loop control begins eliminating the deviation. 

We investigate three aspects which have significant impact on the deviation: (i) the last received command before a gap, because it determines the trajectory extrapolation method used by the local controller during the gap, (ii) how well the shapes of the extrapolated movement and the planned trajectory fit together, and (iii) how fast the robot is moving, because the speed has significantly impact on the magnitude of the deviation. 
Next section summarizes the proposed methods, where each method targets one aspect listed above.

\subsection*{Summary of the proposed methods}

The first method (\emph{Method-A}) selects the optimal movement command type by evaluating the effect of a hypothetical  gap that would happen right after the actual command arrival. 
We select between joint and Cartesian space speed commands, which are supported by the local controller. 

The second method (\emph{Method-B}) can be seen as the AI-based extension of the first method, since this method learns an extrapolation function using a deep neural network (DNN).
This model is used for predicting the values of the missing speed commands and generating new ones on the robot side during gaps. 

The third method (\emph{Method-C}) provides deterministic guarantees of keeping the predefined maximum deviation from the planned trajectory if the maximal duration of the gaps is known. 
This method modifies the speed profile of the trajectory, but the scaled trajectory still follows the same path.

\begin{figure}[ht]
    \centering
    \includegraphics[scale=0.7]{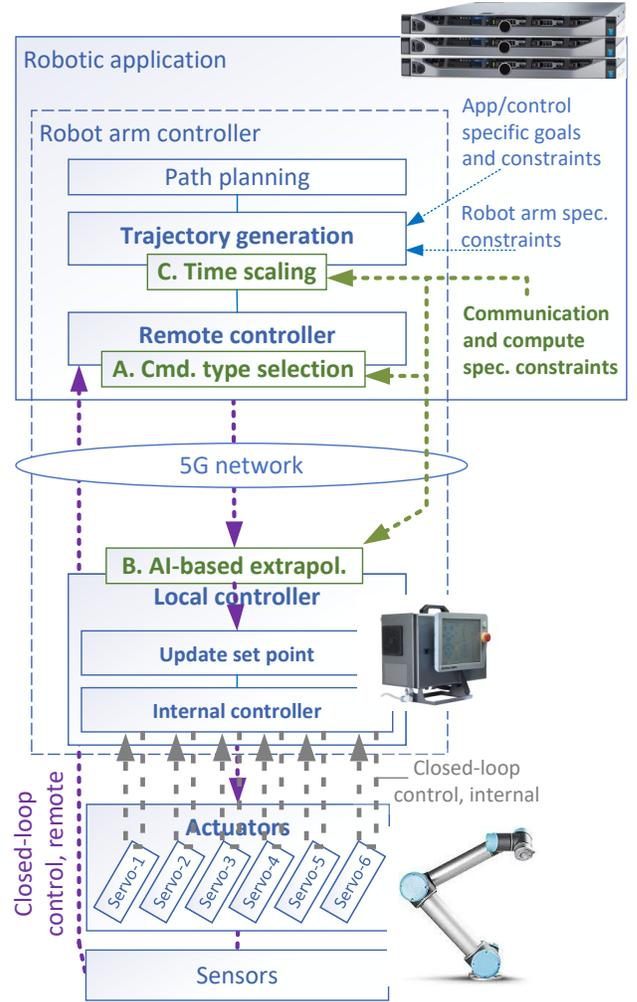}
    \caption{Location of the proposed components}
    \label{fig:components}
\end{figure}

The proposed methods have different implementation requirements and deployment location as Figure \ref{fig:components} shows. \emph{Method-A} can be implemented as part of the remote controller, but the application logic and the implementations on the robot side remain intact. \emph{Method-B} has no impact on remote controller. The DNN can be trained either in the edge cloud or locally at the robot side. The prediction and the artificial command generation are executed on the robot side.
\emph{Method-C} has significant impact on the robotic arm controller application, since it introduces a new trajectory time-scaling functionality which should be integrated into the trajectory generation. 

Let us examine the high-speed industrial robotic welding, where the critical trajectory segment is the periods when the actual welding happens. During this critical segment, any glitch in the motion can easily result in faulty parts, e.g., due to burning through. As the welding speed is increased, it is more critical to be accurate \cite{A3Welding}. 

Translating this example to the notations of Figure \ref{fig:components}, it means that the goal of achieving as high speed as possible with e.g., $<0.5$~mm spatial deviation during welding represents the \emph{App/control specific goals and constraints}. The \emph{Robotic arm spec. constraints} are, for instance, the speed and acceleration profiles of the robot that performs the welding, and the \emph{Communication and compute spec. constraints} are the maximum length of the gaps, e.g., $50$~ms. \emph{Method-C} can provide a scaled trajectory with a new speed profile that satisfies the application specific constraint, i.e., the maximum deviation (accuracy) limit assuming an imperfect communication and compute environment with up to $50$~ms gaps. If the resulting maximum welding speed is not satisfactory from application perspective after scaling, then effort should be focused on providing better communication and compute infrastructure with shorter potential gaps.

\subsection*{Notations}

A pose in Cartesian space (i.e. physical space or task-space) is defined by the $x$, $y$ and $z$ coordinates of the tool of the robotic arm and its orientation is specified by the Euler angles of $r_x$, $r_y$ and $r_z$, we denote by $\underline{p}=\{x,y,z,r_x,r_y,r_z\}$ and by $\underline{p}(t)$ when it is considered as a function of time. The corresponding speed functions denoted by $\underline{v}(t)$.

Let $\underline{q}=\{q_0, q_1 \ldots, q_5\}$ denote the angular joint positions and by $\underline{q}(t)$ when it is considered as a function of time. Similarly, let $\underline{\dot{q}}(t)=\{\dot{q}_0(t), \dot{q}_1(t) \ldots, \dot{q}_5(t)\}$ denote the angular speed of the joints at time $t$. A trajectory $\mathcal{T}$, is uniquely described by joint position functions as $\mathcal{T}=\{ \underline{q}(t), t \in [0,T] \}$ and vice versa.

\subsection{Method-A - Adaptive command type}
\label{sec:methodA}

The idea of the method is that before sending a movement command, we evaluate the gap concealment performance of each potential command type and then we send the best one.

The local controller of an UR5e robotic arm accepts two types of commands that can be used in a velocity control. The joint space speed command is called \texttt{speedj} \cite{URScript} and it accelerates linearly in joint space and continues with constant joint speed. The \texttt{speedl} command behaves similarly in Cartesian space, i.e. it accelerates linearly in Cartesian space and continues with constant tool speed in Cartesian space.

In case of \texttt{speedj}, the angular speeds of the $6$ joints are passed as arguments. In case of \texttt{speedl}, the speed in $x$, $y$ and $z$ directions and the speed of orientation change are passed. 
A speed command is valid and enforced by the local controller until a new speed command arrives or a predefined time-out expires, see Appendix \ref{apendix:speedl} for details. 
We note that similar control interfaces are typically available for other robot vendors as well. Sometimes they require additional modules, custom SDKs/APIs, or other software components to enable per joint control. For instance, similar behavior can be achieved using ABB robots with the Robot Operating System \cite{ABBROSI}.

This method is applicable when the local controller supports more movement command types and can change among them command-by-command basis.
To keep the control loop performance untouched during normal (i.e. no gap) operation,
each potential command type should have the same short-term behavior, i.e. resulting the same set point for internal controller.

\begin{figure}[ht]
    \centering
    \includegraphics[scale=0.46]{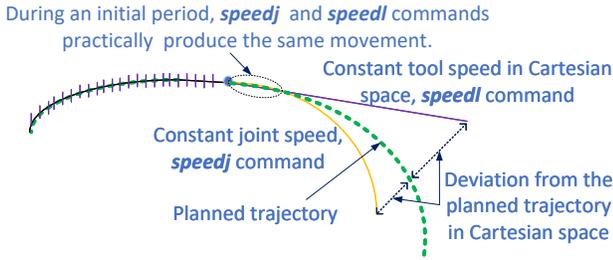}
    \caption{Illustration the robot behavior during gap for \texttt{speedj} and \texttt{speedl} commands. 
    After an initial period, the two trajectories deviate from the planned trajectory following different paths.
    }
    \label{fig:domainIllustration}
\end{figure}

Figure \ref{fig:domainIllustration} illustrates what happens during a gap when velocity control is used. 
The robot continues its movement with constant speed and the path is linear either in joint space or in Cartesian space depending on the type of the last received command. 
After an initial period, the extrapolated trajectory segments result in different deviation from the planned trajectory.

The pseudo-code of Algorithm \ref{alg:getDomain} describes \emph{Method-A} which minimizes the deviation from planned trajectory assuming at most $\Delta$-long gaps and two speed command types. The algorithm also supports the cases where the trajectory is not known in advanced, e.g. visual servoing.

\begin{algorithm}
\small
\caption{Command type selection}\label{alg:getDomain}
\begin{algorithmic}[1]
\Function{getCommand}{$\underline{\dot{q}},~\underline{q},~\Delta,~\mathcal{T},~t$}
\State $\epsilon_1 \gets 2~ms,~~\epsilon_2 \gets 10^{-6},~~f_1=10$

\State \# \texttt{speedl} command's params. by short-term fitting
\State $\underline{v} \gets \mathbf{J}\hspace{-1mm}\left( \underline{q} \right) \cdot \underline{\dot{q}}$ \algorithmiccomment{Speed in Cartesian space}

\State \# Check short-term difference out
\If{$\Vert  \mathbf{FK}\left(\underline{q} + \epsilon_1~\underline{\dot{q}}\right) -  \left( \mathbf{FK}\left(\underline{q} \right) + \epsilon_1~\underline{v} \right) \Vert_2 > \epsilon_2$}
  \State \Return \texttt{speedj} $\dot{q}_0~\dot{q}_1~\dot{q}_2~\dot{q}_3~\dot{q}_4~\dot{q}_5$ 
\EndIf

\State \# Calculate maximal deviation during a gap
\State $D_{J} \gets 0$ \algorithmiccomment{Max. dev. using speedj command} 
\State $D_{C} \gets 0$ \algorithmiccomment{Max. dev. using speedl command} 
\State $\underline{p}_{ref} \gets \mathbf{FK}\left(\underline{q}\right)$\algorithmiccomment{Actual tool pose in Cart.}
\For{$\delta:=0$ To $\Delta$}   \algorithmiccomment{Step size of e.g. $\Delta/10$}
    \State \# Update ref. position if planned traj. available
    \If{$\mathcal{T} \neq \mathbf{empty}$} 
        \State $\underline{p}_{ref} \gets \underline{p}_\mathcal{T}(t + \delta)$ \algorithmiccomment{Planned traj. pose in Cart.}
    \EndIf
    \State \# Determine extrapolated poses
    \State $\underline{\widehat{p}}_{J} \gets \mathbf{FK}\left(\underline{q} + \delta~\underline{\dot{q}}\right)$ 
           \algorithmiccomment{Extrapol. by speedj cmd.}
    \State $\underline{\widehat{p}}_{C} \gets \mathbf{FK}\left(\underline{q} \right) + \delta~\underline{v}$ 
           \algorithmiccomment{Extrapol. by speedl cmd.}
    
    \State \# Avoid unreachable pose and singular region
    \State $\underline{s} \gets \mathbf{IK_{ext}}\left(  \underline{\widehat{p}}_{C}, \underline{q}  \right)$ \algorithmiccomment{Only the closest sol. to $\underline{q}$}

    \If{$\underline{s} = \mathbf{empty}~~\mathbf{or}~~\Vert \underline{s} - \underline{q} \Vert_1 > f_1 \delta \Vert \underline{\dot{q}} \Vert_1$ } 
        \State \Return \texttt{speedj} $\dot{q}_0~\dot{q}_1~\dot{q}_2~\dot{q}_3~\dot{q}_4~\dot{q}_5$
    \EndIf

    \State \# Calculate Cartesian distances   
    \State $D_{J} \gets \mathbf{max}\left[d\left( \underline{p}_{ref},~\underline{\widehat{p}}_{J} \right),~D_{J} \right]$
    \State $D_{C} \gets \mathbf{max}\left[d\left( \underline{p}_{ref},~\underline{\widehat{p}}_{C} \right),~D_{C} \right]$
    \
\EndFor
\State \# Select command with lower deviation
\If{$D_{J} \leq D_{C}$} 
    \State \Return \texttt{speedj} $\dot{q}_0~\dot{q}_1~\dot{q}_2~\dot{q}_3~\dot{q}_4~\dot{q}_5$
\Else
     \State \Return \texttt{speedl} $v_0~v_1~v_2~v_3~v_4~v_5$
\EndIf
\EndFunction
\end{algorithmic}
\end{algorithm}

The algorithm gets the joint speed values ($\underline{\dot{q}}$) calculated by the remote joint velocity controller, the actual position of the joints ($\underline{q}$) available in the remote controller, the maximum considered gap ($\Delta$) and optionally
the planned trajectory under execution ($\mathcal{T}$). 
The deviation is measured to the planned trajectory. If planned trajectory is not available, the deviation is measured to the actual position.

The main steps of the algorithm are the followings.
First, it checks whether the corresponding \texttt{speedl} command would be close enough to the \texttt{speedj} command for the initial period (Line 3-8).
If the \texttt{speedl} command is not close enough, then the \texttt{speedj} command is returned (Line 7).
Then, it determines (Line 18-20) and checks (Line 21-25) extrapolated poses to avoid unreachable pose and singular region (Line 18-25).
Next, the maximal deviations are calculated (Line 26-28).  
Finally, the command with lower deviation is returned (Line 30-35).

The function of $d(\cdot,\cdot)$ calculates the distance between two poses. 
The distance function should be aligned with the goals of the robotic application. 
During evaluation we apply the Euclidean distance, but e.g., orientation error can also be incorporated in the distance function.

To determine the deviation for a \texttt{speedl} command, the joint space velocity is transformed into Cartesian space velocity using the Jacobian matrix $\mathbf{J}\hspace{-1mm}\left( \underline{q} \right)$. 
The forward kinematics function, $\mathbf{FK}$, provides the Cartesian pose from joint position, i.e. $\underline{p}=\textbf{FK}\left( \underline{q} \right)$. 
The forward kinematics function and the Jacobian matrix are robotic arm specific.

\begin{figure}[ht]
    \centering
    \includegraphics[scale=0.45]{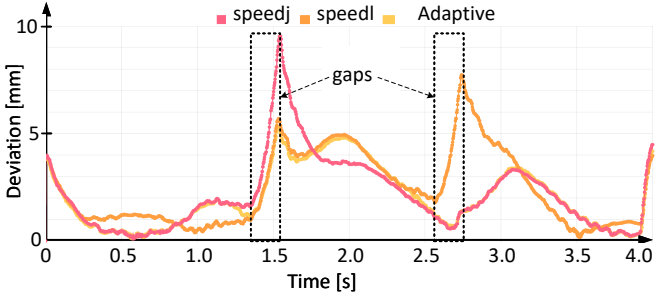}
    \caption{Illustration of the \emph{measured} deviation from the planned trajectory for only \texttt{speedj}, only \texttt{speedl} commands and the adaptive method. The control loop is intentionally not tuned to highlights and enlarge transient periods.}
    \label{fig:domainChangePathDistance}
\end{figure}

We illustrate the operation of this method through measurements done in a private 5G SA network, which connects the local and remote controllers (see in Figure \ref{fig:architecture}). 
To make the operation of the method visible, we added two $200$~ms long artificial gaps, denoted by the dashed rectangles in Figure \ref{fig:domainChangePathDistance}. 
In order to suppress dynamic effects, we did not place any load on the robotic arm. 
To make the transient periods more significant and easier to see, we used a feed-forward proportional remote controller without any compensation technique.

We conducted measurements of the two static methods, i.e., sending only \texttt{speedj} and only \texttt{speedl} commands, respectively. The measurement results of the adaptive method are compared to them.  
Figure \ref{fig:domainChangePathDistance}  shows the {\it measured deviation} of the executed trajectory from the planned trajectory. 

During normal operation, when commands arrive back-to-back, 
the command type should have negligible impact on the operation. 
For $t < 0.25$~s, all of the three methods result in the same deviation. This supports that the command type does not have visible effect.
However, for $t > 0.25$~s, we can observe differences for \texttt{speedj} and \texttt{speedl} commands. This means that the two command types do not have exactly the same short-term behavior, e.g., due to command type specific enforcement of acceleration limit. 
We can also observe that the deviation curve of the adaptive method practically switches between the two reference methods which suggests that the measurement setup provides a dependable environment.

The adaptive method selected \texttt{speedj} command type for $t<1.25$~s and $t>2.3$~s.
For these two segments, the method calculated that a $200$~ms gap would result in lower deviation with joint space linear movement. Despite the actual measure, the deviation is not always lower for that command type. 
For the $[1.25, 2.3]$ interval, the method calculated that Cartesian space linear movement would be better in case of a gap. 

The deviations during the two gaps confirm that the adaptive method selected the command types optimally. 
For the first gap, starting at $t=1.3$~s, the Cartesian space linear movement is performing better, i.e. deviation is about $5.8$~mm instead of $9.8$~mm.
For the second gap, at $t=2.5$~s, the joint space linear movement provides significantly better performance, i.e., $<2$~mm deviation, compared to the deviation of $7.8$~mm for Cartesian space. 

To sum up, the proposed method outperforms the method using only \texttt{speedj} commands about $15$\% in average deviation.
However, more importantly, the method automatically filters poor predictions when it is combined together with the AI-based extrapolation.

\subsection{Method-B - AI-based extrapolation}
\label{sec:methodB}

The idea of the method is to generate artificial command messages to fill gaps in the flow of incoming commands at the robot side. 
These commands extrapolate the trajectory based on 
the history of the current trajectory and the knowledge learnt about the trajectories. 
We use AI models to predict the arguments of these speed command messages.
Without this method, the shape and the speed of the movement are determined by the command received right before the gap. 

We use this method together with the Method-A in order to discard inaccurate extrapolations. 
The AI-based extrapolation is only selected by Method-A when it outperforms the gap concealment performance of the corresponding \texttt{speedj} and \texttt{speedl} commands.

\begin{figure}[ht]
    \centering
    \includegraphics[scale=0.47]{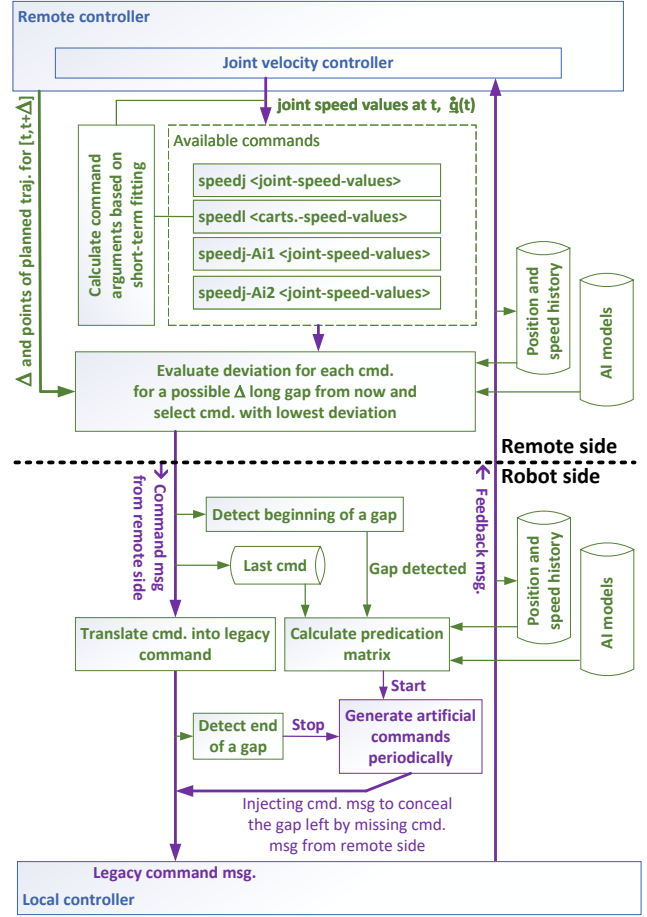}
    \caption{The extended AI-based extrapolation method}
    \label{fig:ab}
\end{figure}

Figure \ref{fig:ab} illustrates the details of this {\it extended method}. 
First, we introduce a new command type (e.g. \texttt{speedj-Ai1}) which behaves like a legacy \texttt{speedj} command, but it activates the AI-based extrapolation at the robot side. It is straightforward to introduce multiple commands which can use different underlying AI models.

For consistent operation of the remote and robot sides, we need to fulfill further requirements. This is needed since the remote side calculation assumes that it knows what commands will be generated at the robot side.
First, the command type selection method (at the remote side) and the AI-based extrapolation method (at the robot side) 
should use the same AI model for evaluation and prediction.
Second, the input of the AI model during the evaluation and the prediction should be the same. To satisfy the first requirement, we need to use a synchronized solution for propagating the trained AI model between the remote and local sides. To satisfy the second requirement, we should use the position and speed values that are already available at both sides. 
The historical position information originates in the robot side and is conveyed by feedback messages to the remote side. 
To consider propagation delays we can set $\alpha_1> D_{feedback}$ in (\ref{eq:HO}), where $D_{feedback}$ is the maximum delay of the feedback messages.

At the remote side, this new command is also considered in the command type selection method.
Since the new command will be translated into legacy \texttt{speedj} command at robot side, its short-term behavior is the same as of \texttt{speedj}. 
Regarding the max. deviation during a gap, we use the extrapolated speed values generated by the AI-based extrapolation method, and to get $\underline{\widehat{p}}$ values we integrate the extrapolated speed functions.

At the robot side, the introduced command is translated into a legacy \texttt{speedj} command and forwarded to the local controller.
However, if a gap is detected and the introduced command was the latest received one, then the artificial command generation is activated to generate \texttt{speedj} commands until the end of the gap. 

The DNN used in the evaluation takes $N$ joint positions and speed points from the current trajectory as history and predicts $M$ joint speed points. 
The input values can be compiled into a matrix, denoted by $H(t)$ in (\ref{eq:HO}), where $\alpha_i$ denotes the offsets of the $N$ sampling points.  
In the quantitative evaluation we use $N=13$ time points up to $4$~s, which are roughly evenly spaced in log-space. 
\begin{equation}
\small
  H(t) = 
\left[
\begin{array}{c;{2pt/2pt}c}
    \underline{p}(t-\alpha_1)  &  \underline{v}(t-\alpha_1) \\ 
    \underline{p}(t-\alpha_2)  &  \underline{v}(t-\alpha_2) \\ 
    \vdots & \vdots \\ 
    \underline{p}(t-\alpha_{N})  &  \underline{v}(t-\alpha_{N}) \\  
\end{array}
\right]  
~
O(t) = 
\left[
\begin{array}{c}
    \underline{\widehat{v}}^{t}(t+\beta_1)  \\ 
    \underline{\widehat{v}}^{t}(t+\beta_2)   \\ 
    \vdots  \\ 
    \underline{\widehat{v}}^{t}(t+\beta_M)  \\  
\end{array}
\right]
\label{eq:HO}
\end{equation}
The DNN provides the prediction matrix, $O(t) = \mathbf{DNN}( H(t) )$ in (\ref{eq:HO}),
where $\beta_i$ denotes the time points of the prediction and the $\underline{\widehat{v}}^{t}(t)$ denotes the predicted speed values.
The $t$ in superscript indicates that the prediction is calculated based on a $H$ matrix sampled at time $t$.
For predicting the speed values at an intermediate time point, linear interpolation is applied. In the evaluation we use $M=11$ evenly spaced time points between $0$ and $200$~ms. This method allows us to predict joint speed values for the period of $[t+\beta_1, t+\beta_M]$ using position and speed values from the period of $[t-\alpha_N, t-\alpha_1]$. 
For gaps larger than $\beta_M$, the $O$ matrix should be calculated again using also predicted values in its input matrix $H$. 
If the $\alpha_N$ long trajectory history is not available yet, then trajectory staring position and zero speed values are used in the matrix $H$ for missing values. 

In order to further improve performance, ensemble techniques can be applied, 
i.e. extrapolating by merging predictions from multiple models. 
The models can have different architecture, not limiting to the DNNs. In addition, through the proper assortment of training trajectories, we can focus on diverse aspects. For example, dataset that contains (i) trajectories used by the robotic application, represents task specific aspects; (ii) randomly generated trajectories mean robotic arm specific aspects; or (iii) trajectories that were identified as challenging ones can represent the performance specific aspects. 
The evaluation of this technique is beyond the scope of this paper, we used randomly parameterized trajectories as dataset 
to train our DNN model.


\subsection{Method C - Trajectory time-scaling for bounded deviation}
\label{sec:methodC}

The essence of the method is that if the deviation is too large then we temporarily slow down the movement.
We apply trajectory time-scaling to avoid going through accuracy sensitive segments with too high speed. 

This method is feasible for tasks which tolerate trajectory slowdowns.
If the shape of the speed profile is still relevant, then we slow down the trajectory with a constant speed-factor, that we call as {\it static scaling}. Otherwise, the method optimizes the shape of the speed profile as well which is referred to as {\it varying scaling}.

We introduce a supplementary time-scaling function, $s(t)$ or $s$, for a trajectory. This function is defined for the interval $[0,T_b]$, where $T_b$ is the duration of the trajectory. 
The time-scaling function defines a \emph{scaled trajectory} by changing the timing of the \emph{base trajectory}. 
This means that the scaled trajectory goes through the same path as the base one, but its time schedule is different. The joint position and speed functions of the scaled trajectory can be expressed as the functions of corresponding values of the base trajectory and the time-scaling function:
\begin{equation}
    \underline{q}_{s}(t)  = \underline{q}_{b}\left(s(t)\right),~~
    \underline{\dot{q}}_{s}(t)  =  \dot{s}(t)~\underline{\dot{q}}_{b}\left(s(t)\right),~~
    t \in [0, s^{-1}(T_b)]
    \label{eq:scaledTrajectory}
\end{equation}
The function $\dot{s}(t)$, which is the corresponding {\it speed-scaling function}, is the time derivative of $s(t)$.
We restrict our attention to down-scaling of the speed, i.e. we assume $0 < \dot{s}(t) \leq 1$.
The $s^{-1}$ denotes the inverse of $s$ and $T_s = s^{-1}(T_b)$ is the duration of the scaled trajectory.

The proper selection of the scaling function ensures that the maximal deviation during the execution of the scaled trajectory always remains below a predefined limit $L$ even if at most $\Delta$ long gaps occur.
The goal is to (i) provide deterministic accuracy guarantees for the critical segment, which is identified by the robotic application, of the base trajectory meanwhile (ii) apply as minimal as possible trajectory lengthening and (iii) remain below the speed and acceleration limits of the base trajectory. 

We formulate the above goal as a constrained optimization. We try to find an optimal time-scaling function $s^*$ to minimize the duration of the scaled trajectory, i.e,
\begin{equation}
s^* = \argmin_{s} s^{-1}(T_b)
\label{eq:methodC_argmin}
\end{equation}
where the scaled trajectory, which is defined by the $s^*$ according to (\ref{eq:scaledTrajectory}),  fulfills the \emph{constraints} below :
\begin{enumerate}[A.]
    \item $D(\dot{s}, \Delta, T_A, T_B) \triangleq  \\
    \max\limits_{\substack{\delta \leq \Delta \\ t \in \mathcal{C}}}  
    d\left[\mathbf{FK}\left( \underline{q}_{s}\left(t + \delta\right) \right), 
    \mathbf{FK}\left( \underline{\widehat{q}}_{s}^{t}\left(t + \delta\right) \right) \right] \leq L$, \\
            where $\mathcal{C} = \{ t:~s^{-1}(T_A) \leq t \leq s^{-1}(T_B)\}$. 
    \item $\max\limits_{0 \leq t \leq T_s}\left\vert\left\vert \underline{\ddot{q}}_{s}(t) \right\vert\right\vert_\infty \leq \max\limits_{0 \leq t \leq T_b} \left\vert\left\vert \underline{\ddot{q}}_{b}(t) \right\vert\right\vert_\infty$
    \item $\max\limits_{0 \leq t \leq T_s} \left\vert\left\vert \underline{\dot{q}}_{s}(t) \right\vert\right\vert_\infty  \leq \max\limits_{0 \leq t \leq T_b} \left\vert\left\vert \underline{\dot{q}}_{b}(t) \right\vert\right\vert_\infty$
\end{enumerate}

The function $d(\cdot,\cdot)$ denotes the distance function which measures the deviation between two poses, e.g., Euclidean distance.
$T_A$ and $T_B$ denote the start and end time of the critical segment of the base trajectory.
The $\underline{\widehat{q}}_{s}^{t}\left(t + \delta\right)$ denotes the predicted joint position 
at $t+\delta$ when the prediction started at $t$. 

Constraint-A keeps the deviation caused by a gap of maximum size of $\Delta$ below a predefined limit of $L$. The supplementary constraints B and C ensure that the scaled trajectory does not exceed the maximum speed and acceleration of the base trajectory.

\begin{figure}[ht]
    \centering
    \includegraphics[scale=0.58]{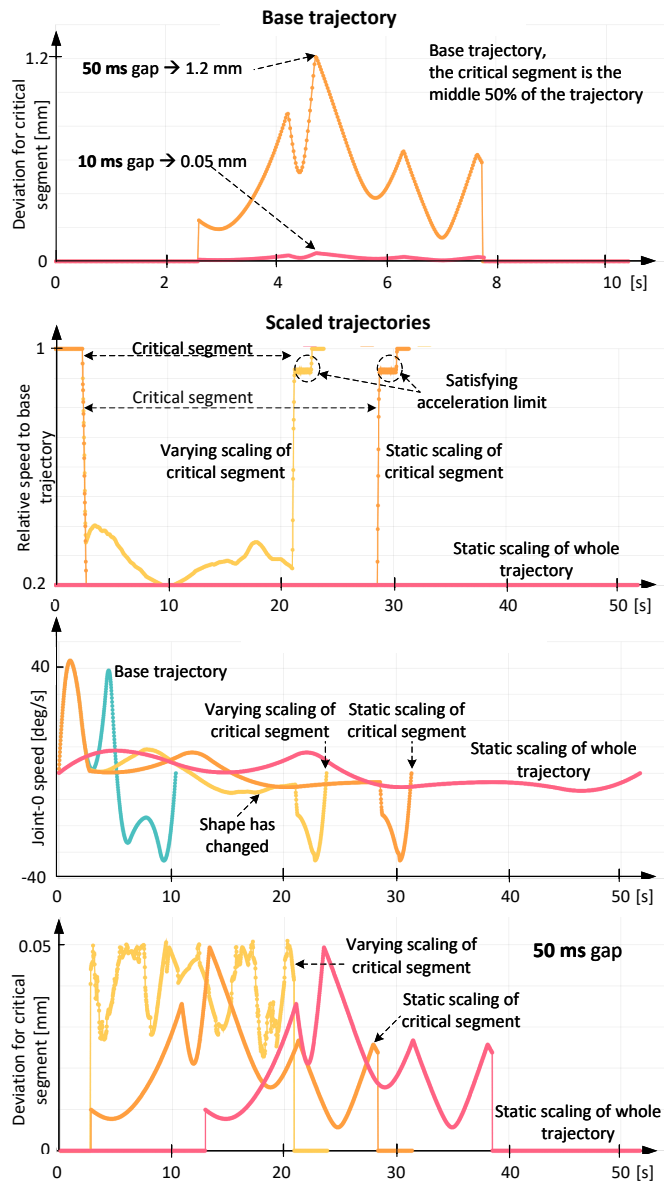}
    \caption{
     {\bf Base trajectory}: deviation from the planned trajectory during $10$ and $50$ ms gaps.
     {\bf Scaled trajectories for $\mathbf{L=0.05}$ mm and $\mathbf{\Delta=50}$ ms}: (i) relative speed functions, 
     (ii) angular speeds of the first servo of the robotic arm,
     (iii) deviation functions for $50$ ms gap.
     }
    \label{fig:s0}
\end{figure}

Figures \ref{fig:s0} provides an example. We have a $10.4$~s long base trajectory. 
Assume that the robotic task tolerates only $L=0.05$ mm deviation during the critical segment ($2.6$ s $\leq t \leq 7.8$ s).
Evaluation shows that the system can fulfill this requirement by a connection with maximum $\Delta = 10$~ms gaps,
However, we would like to use a connection which can guarantee only $\Delta = 50$~ms upper limit on gaps.
If the robotic task tolerates the modification of the trajectory timing, then we can use one of the proposed scaling methods. 

From robotic task point of view, the cost of the scaling method is the potentially longer trajectory duration. 
With static scaling, the required speed-scaling factor is $0.2$ and the length of the scaled trajectories are $52$~s and $31.6$~s, respectively.
With varying scaling, the length gets shorter, i.e., $24$~s, but the shape of joint speed profile has changed. 
Note that, the steps in relative speed function after the critical segment are needed to satisfy the joint acceleration limits.

In the quantitative evaluation, we solve the optimization by a random search method which quickly finds good solutions for our ill-structured constrained optimization problem. 
We approximate the speed-scaling function $\dot{s}$ by a piece-wise linear continues function with $2$~ms resolution, 
which is aligned with the update time of the considered robotic arm. 
To get the time-scaling function we use integration, i.e. $s(t) = \int_0^t \dot{s}(\tau) d\tau$ and $s(0) = 0$.

The following \emph{numerical method for scaling} provides a sub-optimal solution for $\dot{s}(t)$:
\begin{enumerate}[I.]
    \item Find a constant speed-scaling function for which all the three constraints are satisfied on the critical segment. \\
     $\dot{s}(t) = \left\{ c,~\text{if } T_A \leq t \leq T_B;~~1,~\text{otherwise}\right\}$.
    \item Refine speed-scaling function (only for varying scaling). 
    On the critical segment, try to increase the speed-scaling function locally, but keep the three constraints satisfied. 
    Center point of the increase is randomly chosen, and the increase is dispensed to a medium-length, e.g. $500$~ms interval. Too short interval results in a sudden increase in acceleration, too long interval excludes potential solutions. The size of the increase is gradually reduced during the search. 
    This step is repeated until a sub-optimal point is reached.
    \item Update speed-scaling function on the non-critical segments ($t<T_A$ or $t>T_B$) such 
    that the maximal speed and acceleration limits are kept for that part.
    \item Make speed-scaling function feasible. The speed-scaling function on the whole domain is gradually reduced until all constraints are satisfied on the whole domain.
\end{enumerate}

\subsection{Considerations}

However, every proposed method uses the maximum gap duration, $\Delta$, as input, it is used differently and in this way the accuracy requirements on $\Delta$ are also different. The performance guarantees that are provided by the trajectory time-scaling method are valid if the duration of a gap does not exceed the $\Delta$. 
For longer gaps, the performance guarantees could be violated. 
In the other methods, the $\Delta$ plays less critical role since these methods aim to improve performance in average and the $\Delta$ gives a target time horizon for which the performance optimizations are focused.

In the time-scaling method, we assume that the planned trajectory is realized accurately by the robotic arm in normal operation. 
Manufactures of industrial robotic arms are improving robot accuracy and repeatability, which spans from improvement of mechanical components like gearbox or bearings to using advanced compensation models in the controllers \cite{RobotAccuracy}, \cite{RobotAccuracy2}.
These efforts support that a planned trajectory can be realized with negligible error in normal operation,
In this way the performance guarantees for gaps remain valid during execution. 
If the robot inaccuracy is not negligible, then this type of imperfection should also be incorporated into the method, e.g. by including as an additional margin in the deviation target. 

\section{Quantitative evaluation
\label{sec:quantitative_evaluation}} 

\subsection{Evaluation methodology and trajectories used for evaluation}

We used three environments depending on the purpose of the evaluation as detailed in Appendix \ref{appendix:eval_methodology}.

We created trajectories of each with $9$ randomly generated waypoints between two surfaces in the task-space of the UR5e robot%
\footnote{
The surfaces defined as: $x=\pm 0.4$, $0.1 \leq y \leq 0.7$ and $-0.3 \leq z \leq 0.6$
}. 
The trajectories travel through these waypoints and cubic spline interpolation is used among waypoints. 
Regarding the timing of the trajectories, we generated as short as possible trajectories while considering maximum speed and acceleration limits coming from the constraints of the robot%
\footnote{Maximum joint speed is set to $1/4$ of the maximal robot speed.
}.
The trajectories were filtered to contain only those ones that can be realized by the UR5e robot and thus we ended up with $4000$ such trajectories that we also serialized and archived for the purpose of reproducibility.

\subsection{Evaluation of the AI-based extrapolation}

The method is described briefly in Section \ref{sec:methodB} and aims to replace missing control commands at the robot side with artificially generated commands based on predictions of a trained DNN model. 
We limit our attention to assess what magnitude of gain can be achieved and we do not target to optimize the model itself. 

We evaluated the performance of the DNNs which use $13$ historical time points and predict $11$ joint speed points. 
The historical time points, i.e. $\alpha$ values in (\ref{eq:HO}), are $0, 0.05, 0.1, 0.17, 0.25, 0.37, 0.5, 0.75, 1, 1.5, 2, 3, 4$~s. The time points of prediction are equally spaced  between $0$ and $200$ ms.
The input size of the DNN is $156$ consisting of joint position and speed values, the output size is $66$ consisting of predicted joint speed values. Hyperbolic tangent activation functions are used. The input position/speed values are mapped in the $(-1,1)$ interval and for the predicted speed values the inverse transformation is applied. 

We apply supervised learning method to optimize the parameters of the neural network. We used $3800$ trajectories for training and the remaining $200$ for validation. We randomly sampled $32768$ segments from the training trajectories. These $4.2$~s long segments were split into two parts and used as labeled data. The first $4$~s was the input part and the final $0.2$~s was the label.

\begin{figure}[ht]
    \centering
    \includegraphics[scale=0.46]{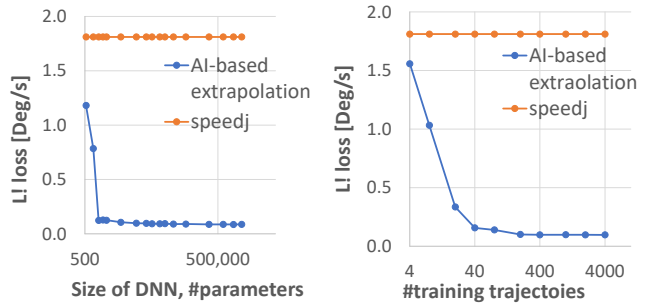}
    \caption{The L1 loss versus the model size and the number of trajectories used during the training. 
    As reference the constant joint speed extrapolation method of \texttt{speedj} command is used.}
    \label{fig:dnnPar}
\end{figure}

Figure \ref{fig:dnnPar}  shows the error of the trained network as the function of the model size and the number of training trajectories, respectively. We can observe around $10$x smaller $L1$ loss for DNNs with a few thousands parameters already, than with the reference method. This means that relatively small model is suitable to harvest an order of magnitude gain, which is beneficial, since a small model can enable fast training and prediction times, as well as there is no need for dedicated HW for execution on the robot side, i.e., the local controller of the robot may be reused to execute the prediction.
We can also observe that ${\sim}40$ trajectories are already providing highly accurate predictions. Another conclusion is that there is no need to use more than a few thousands of trajectories for the training since the precision is not improved considerably. 

\begin{table}[ht]
    \centering
    \begin{tabular}{c|ccccccc}
         \hline
         \#layers & 2      & 3     &   4   & 5     & 7     & 10    & 25 \\
         \hline
          Loss        & 0.098  & 0.097 & 0.096 & 0.098 & 0.100 & 0.104 & 0.110 \\
         \hline  
    \end{tabular}
    \vspace{2mm}
    \caption{L1 loss [deg/s] versus number of hidden layers}
    \label{tab:layers}
\end{table}

In Table \ref{tab:layers} we show the relation between the internal structure of the DNN, more specifically the number of hidden layers and the L1 loss of the prediction. We set the number of internal parameters to about $50000$ and adjusted the number of hidden layers and the width of the layers (i.e., number of neurons per layer) accordingly. We can conclude that the accuracy of the model basically independent of the architecture of the network, i.e., there is no significant difference in the prediction accuracy as the architecture of the DNN is changed. Furthermore, numbers indicate that few hidden layers, $3$ or $4$, are enough to be used, no need to implement more complex structures. 
During training, we did not observe overfitting and the training process seemed to be robust against training trajectories and parameter setting. 

We conclude that DNNs can be effectively applied for learning the trajectories and using it for predicting missing control commands during gaps. The accuracy outperforms the reference solution by an order of magnitude even in this non-optimized implementation. To achieve this performance, we only need a small DNN with relatively small training dataset which enables us to train the model quickly. For the further numerical evaluations, we apply $3$ hidden layers with $112$ neurons and about 50k parameters.

\subsection{Evaluation of the trajectory time-scaling method together with the extended AI-based extrapolation method}

The \emph{trajectory time-scaling} method is explained in detail in Section \ref{sec:methodC}. 
Its goal is to adjust the timing of the trajectory in advance to ensure that the maximum deviation from the planned trajectory would always be kept within the predefined limit.

\begin{figure}[ht!]
    \centering
    \includegraphics[scale=0.47]{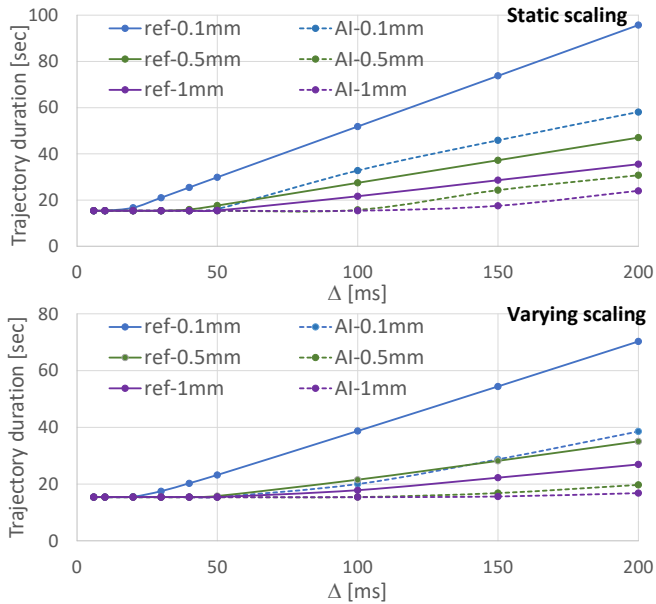}
    \caption{
    The average duration of $200$ scaled trajectories is illustrated as a function of gap duration, $\Delta$.
    Solid lines correspond to reference method where \texttt{speedj} commands are used.
    Dashed curves shows the results when the extended method, i.e. AI-based extrapolation, is used.
    The methods are evaluate for different accuracy limits, $L$. 
    }
    \label{fig:scale}
\end{figure}

In Figure \ref{fig:scale}, we depict the average duration of the scaled trajectories as a function of maximum gap duration, $\Delta$.
On the remote controller side, we evaluate deviations for AI-based extrapolation, \texttt{speedj} and \texttt{speedl} commands and send the command with the lowest deviation. 
On the robot side, we generate artificial commands if needed. 
These results are labeled as \emph{AI} in the figure and marked by dashed lines.
Moreover, we carried out evaluations for different predefined accuracy limits $L$ such as $0.1, 0.5$ and $1$~mm. 
We also use constant joint speed commands (last received ones) to replace missing commands, this is referred to as \emph{ref} and marked by solid lines. 

We can extract the required trajectory lengthening for $L$ and $\Delta$ values from the Figure \ref{fig:scale}.
For instance, if we would like to guarantee e.g., $L=1$~mm accuracy during maximum e.g., $\Delta=200$~ms long gaps, we need to lengthen the trajectories with static scaling method from ${\sim}16$~s to ${\sim}36$~s in average if constant joint speeds are used (ref-$1$~mm, solid purple curve). While if we apply the extended method (AI-$1$~mm, dashed purple line), then we need to lengthen the trajectories in average only from ${\sim}16$~s to ${\sim}23$~s. If we check the same curve at $100$~ms, we can conclude that with the baseline case (ref-$1$~mm at $100$~ms), the trajectory duration needs to be extended from ${\sim}16$~s to ${\sim}21$~s to keep the $1$~mm accuracy, while with the extended method (AI-$1$~mm at $100$~ms) we do not need to touch the trajectory duration at all and we can still ensure the $1$~mm accuracy. The numbers in Figure \ref{fig:scale} are averaged over $200$ trajectories. Moreover, we define a critical segment for each trajectory which is $50\%$ of the trajectory, i.e., it is between $T_A=0.25\times$trajectory duration and $T_B=0.75\times$trajectory duration.

We can also see that the varying scaling method further decreases the duration of scaled trajectory. For instance, let's take the same purple curves, in order to ensure e.g., $1$~mm accuracy during maximum e.g., $\Delta=200$~ms long gaps, we need to lengthen the trajectories from ${\sim}16$~s to ${\sim}27$~s (instead of ${\sim}36$~s as with static scaling in Fig. \ref{fig:scale}) in average if constant joint speeds are used (ref-$1$~mm, solid purple curve). 
While in case of extended method (AI-$1$~mm, dashed purple line), we need to lengthen the trajectories in average only from ${\sim}16$~s to ${\sim}17$~s (compared to ${\sim}23$~s as with static scaling).

\begin{figure}[ht]
    \centering
    \includegraphics[scale=0.45]{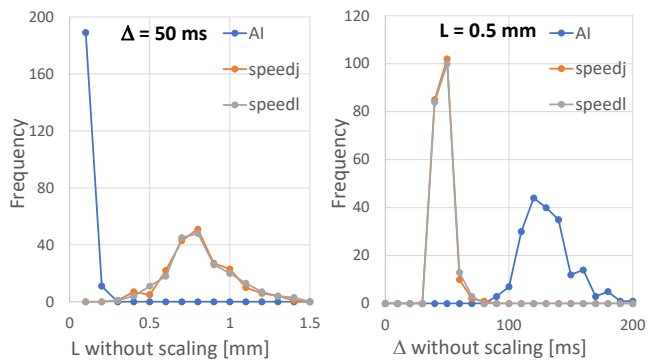}
    \caption{Maximal deviation and gap without need for scaling. Histogram for the $200$ validation trajectories, bin size is $0.1$ mm and $10$ ms, respectively.}
    \label{fig:maxDelta}
\end{figure}

Figure \ref{fig:maxDelta} shows the histogram of maximal deviation (left) and gap duration (right) that are supported without scaling. 
As we expected, the AI-base extrapolation method improves performance significantly. The reference methods perform very similarly to each other. To achieve higher accuracy, i.e. lower $L$, or better gap tolerance, i.e. higher $\Delta$, needs scaling. 

\begin{figure}[ht]
    \centering
    \includegraphics[scale=0.45]{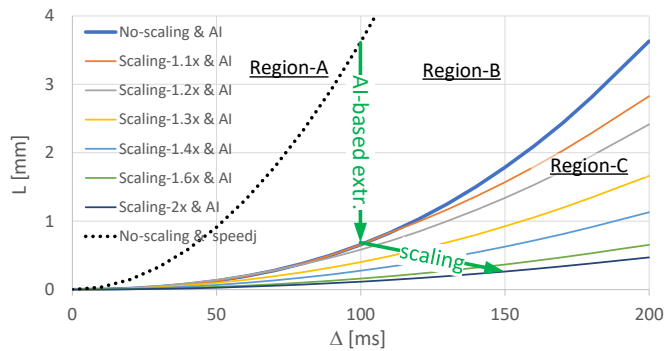}
    \caption{The $L-\Delta$ space for a particular trajectory. Static scaling. 
    Region-A: feasible points with \texttt{speedj} commands.
    Region-B: the AI-based extrapolation method makes these points also feasible. 
    Region-C: for points of this region trajectory scaling is needed.
    } 
    \label{fig:regio}
\end{figure}

Figure \ref{fig:regio} illustrates the regions in the $L$-$\Delta$ space of a trajectory. Points of Region-A are feasible without AI-based extrapolation and scaling. Points of Region-B are feasible without scaling only if AI-based extrapolation is used. Points of Region-C require AI-based extrapolation and also trajectory scaling. For example, $L=0.25$ mm maximal deviation with $\Delta=150$ ms max. gaps can be achieved using (i) AI-based extrapolation and also (ii) trajectory scaling where the scaling method doubles the trajectory duration. 

\begin{table}[ht]
    \centering
    \begin{tabular}{|c|ccc|}
    \hline     
         $\Delta$ & speedj & speedl & speedj-AI\\
         \hline
         \hline
         50ms	& 0.85\% & 0.87\% & 98.28\%  \\
         100ms	& 1.46\% & 1.49\% & 97.04\%  \\ 
         150ms	& 1.99\% & 1.57\% & 96.44\%  \\
         200ms	& 3.59\% & 2.21\% & 94.20\%	 \\
         \hline
         Avg     & 1.98\% & 1.53\% & 96.49\% \\
         \hline
    \end{tabular}
    \vspace{2mm}
    \caption{Fraction of command types, Static scaling and $L=0.5$ mm.}
    \label{tab:frac}
\end{table}

Table \ref{tab:frac} provides the fraction of different command types. 
Most of the commands were our new command for AI-based extrapolation and 
the reaming commands were \texttt{speedj} and \texttt{speedl}. 
In this way, inaccurate predictions ($3$-$4$\%) of AI-based extrapolation were hidden by sending legacy commands.
This mechanism allows us to relax the reliability requirement on the AI-based extrapolation method.
  
To sum up, the quantitative evaluation illustrates that describing the imperfection of the underlying components as gaps already allows us to achieve significant performance gain. 
Meanwhile, we believe, this approach keeps the cost of the implementation reasonable. 
Each underlying component, like wireless network and cloud execution environment, only needs to figure out its worst-case performance and communicates it to the methods as maximal gap duration. We proposed methods which use this information as input and provide performance improvement realized at robotic application level. 
The main advantage of the proposed methods is that (i) the methods can be combined with each other, (ii) the gains add together and (iii) culminates in robotic application-level gain. 

\section{Conclusions}

This paper proposes different performance optimization methods leveraging the concept of communication-compute-control co-design by taking into account some characteristics of the communication and compute technologies into the control application design and implementation. 
Such co-design is particularly relevant for two main trends in cyber-physical systems: 
(i) the adoption of cloud computing for efficient and scalable high-performance computing, as well as 
(ii) the introduction of wireless networking that provides flexibility in system deployment.

The trend of targeting lower determinism together with higher interworking facilitates the convergence of communication, compute and application.
We have shown that co-design is a promising approach, if the time-characteristics of the compute (i.e. computation time for the control command) or the communication (transferring the control command and/or sensor feedback between the controller and the robot) are subject stochastic variations. 
We assumed that any such variations can be modelled by the control application as ``gaps''. 
We have proposed and analyzed co-designed control logic, where the control application anticipates and compensates possible imperfections of the compute and communication systems. 
From compute and communication perspective, it is expected that those systems can estimate and quantify their imperfections in terms of a maximum gap, that is then considered by the control algorithm for robust control design.

We have implemented these methods and their combination in a robotic use case where a UR5e robotic arm is controlled from an edge cloud environment and connected to the robot over a 5G network. 
We have demonstrated that a low-complexity deep neural network-based command loss concealment solution can reach an order of magnitude trajectory extrapolation accuracy improvement. 
In addition, this performance gain can be turned into significant gains in the trajectory time-scaling method resulting in up to 45\% shorter trajectories. 
To conclude, the co-design-based methods have considerable potential to compensate for the performance imperfections of the environment from the industrial application point of view.

\section*{Acknowledgements}
This work was supported by the European Union’s Horizon 2020 research and innovation programme through DETERMINISTIC6G project under Grant Agreement no. 101096504.

\appendix

\subsection{Explorative measurements and the selected baseline system}
\label{appendix:background}

This section describes how we have selected our baseline system to which the performance of the proposed methods was compared.
First, we did explorative measurements in various setups for PLC-based solutions as part of a collaboration project briefly discussed in \cite{Comau}. After concluding the results, we built up our baseline system. 

In the measurements, the SoftPLC, which is a software-based version of the PLC, executes a robotic application in the edge cloud and is connected to the local controllers of a robotic arm. 
This solution can already utilize the advantage of cloud execution and wireless connection and requires minimal modification of the original setup. 
However, the downside is that the cloud and the wireless connection introduce additional delay/jitter into the communication between the PLC and local controller.
Industrial protocols, such as PROFINET \cite{PROFINET}, typically tolerates increased delay, but are more sensitive to jitter. It tolerates only a few communication cycles, typically $3$, without valid data reception before a communication failure is reported. 
In order to handle increased jitter, we need to configure the update times of the industrial protocols so that they can tolerate the largest acceptable jitter.
We also need to re-optimize the controllers to compensate the introduced delay.
To understand the technical consequences of the introduction of cloud and wireless in a PLC-based system, we conducted some explorative measurements in different setups.

\begin{figure}[ht]
    \centering
    \includegraphics[scale=0.53]{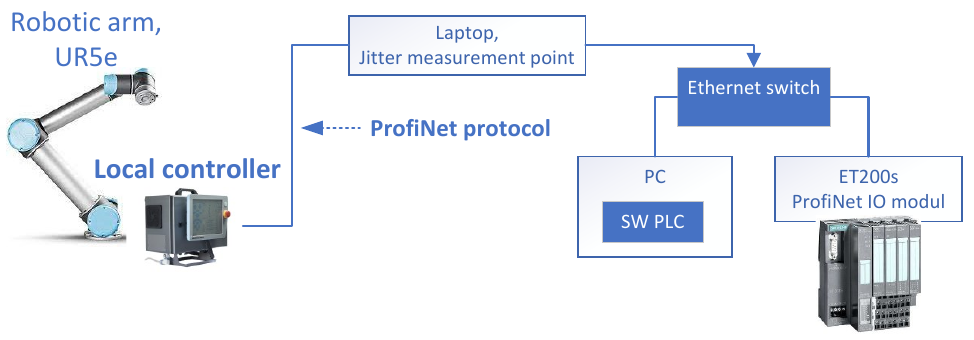}
    \caption{Measurement setup to determine the contribution of cloud environment to communication jitter}
    \label{fig:m_cloud}
\end{figure}

First, we started by measuring the impact of the cloud environment on communication jitter. 
We investigated a system using (hardware) PLC as the baseline and also systems using the Siemens SIMATIC WinAC RTX F 2010 SoftPLC \cite{SoftPLC} in different virtualization solutions.
Figure \ref{fig:m_cloud} shows the measurement setup. We measured the per packet jitter (standard deviation of packet inter-arrival times) of PROFINET Real Time Class 1 frames.

Some relevant measurement results are listed below:
\begin{itemize}
  \item Hardware PLC, stable operation: 36.6~us jitter
  \item SoftPLC on purpose-specific hardware, stable operation: 23.4~us jitter
  \item SoftPLC on generic hardware, bare metal hypervisor, stable operation: 286~us jitter
  \item SoftPLC on generic hardware with KVM on Ubuntu with PCI pass-through, stable operation: 279~us jitter
  \item SoftPLC on generic hardware with KVM on Ubuntu with e1000e network driver emulation, {\it unstable operation}: 293~us jitter
\end{itemize}

We used Dell PowerEdge R420 servers as generic hardware. 
PCI pass-through allows the SoftPLC to use the entire physical Network Interface Card (NIC), which has the consequence that the NIC is not available to other SoftPLC anymore. 
A NIC can be shared among more SoftPLCs with driver emulation, but during a one-hour measurement, we observed unstable operation. The controller stopped operating occasionally and the real-time OS under the SoftPLC (IntervalZero \cite{IntervalZero}) had to be restarted. Single root I/O virtualization (SR-IOV) was not supported by any of the NICs suitable for this SoftPLC. 
We concluded that it is challenging (i) to configure and run SoftPLCs in edge cloud, as well as (ii) to share the hardware resources among them.

\begin{figure}[ht]
    \centering
    \includegraphics[scale=0.51]{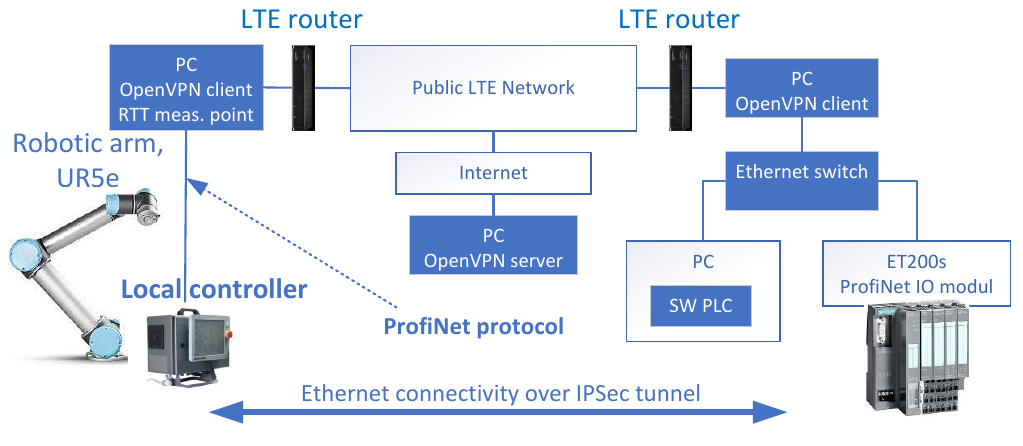}
    \caption{Measurement setup to understand the contribution of wireless connection to communication jitter and delay}
    \label{fig:m_LTE}
\end{figure}

We also measured the contribution of wireless connections to the communication delay. We used the publicly available LTE Mobile Broadband service of a Hungarian operator. Figure \ref{fig:m_LTE} illustrates the measurement setup.

\begin{figure}[ht]
    \centering
    \includegraphics[scale=0.54]{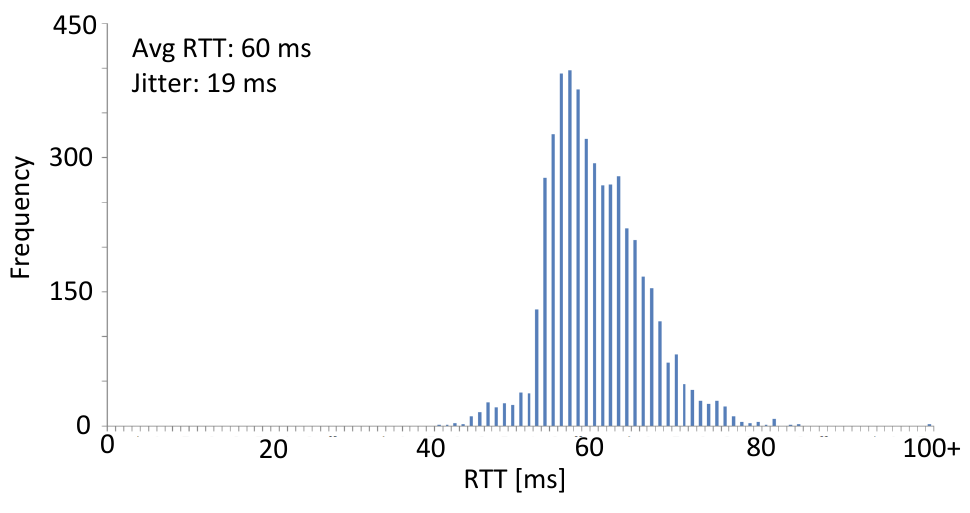}
    \caption{Histogram of measured RTT over a public LTE network. $3$ samples in the bin of $\geq 100$ ms. 
    Two LTE connections are involved in the RTT.
    }
    \label{fig:m_jitter}
\end{figure}

We depict the measured round-trip-time (RTT) in Figure \ref{fig:m_jitter}. 
During the $10$-minute-long measurement the PROFINET connection reported connection failures (the gap was larger than $3 \times 16$ ms) $3$ times. 
For an update time less than $16$ ms, we observed that the PROFINET connection could not be established.

To sum up, PLC-based solutions are sensitive to the glitches of cloud environment and the jitter from wireless connection. We also concluded that stable operation of multi-user scenarios could be more complex, e.g. (i) running more SoftPLCs on the same generic hardware, (ii) sharing the NIC, or (iii) serving devices by a loaded radio cell. 

As a conclusion of the above experimental architectures and measurements, we defined our baseline architecture,
which is depicted in Fig.~\ref{fig:architecture}. We replaced the PROFINET protocol with a TCP/IP based, also periodic protocol (URScript \cite{URScript}) to keep the same traffic pattern.
Furthermore, we implemented a custom robotic arm controller instead of using the SoftPLC that allows us to better intervene into the functions of the robot controller to implement co-design related performance improvement methods. 
Thus, we intended to have a system with similar behavior as the system illustrated in Fig. \ref{fig:m_LTE} 
while still having the freedom to modify the application as needed.

\subsection{Measuring the quality of communication between remote and local controller}
\label{appendix:LTE_RTT}

\begin{figure}[h]
    \centering
    \includegraphics[scale=0.49]{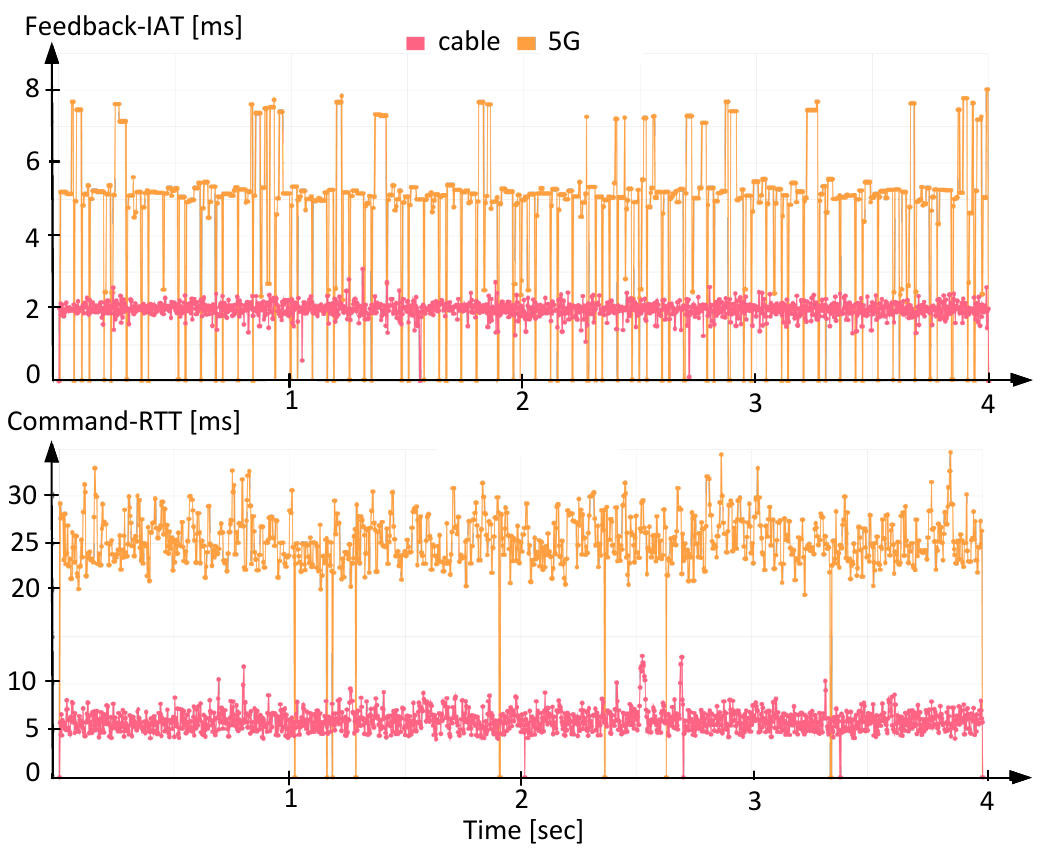}
    \caption{IAT of feedback messages and RTT of \texttt{speedj} command messages for a $4$-sec-long trajectory over cable and 5G. Zero values in the RTT chart only means that the measurement was skipped.}
    \label{fig:m_iat_rtt}
\end{figure}

Figure \ref{fig:m_iat_rtt} shows the measured inter-arrival time (IAT) of the feedback messages and the RTT of \texttt{speedj} command messages for a $4$~s long trajectory in our baseline architecture. 

 \subsubsection{Feedback-IAT} The local controller sends feedback messages with $2$~ms period in uplink direction. The IAT of the feedback messages are measured in the remote controller. 
 We observed a small jitter for cable, which is caused by our baseline system excluding wireless part. 
 The local controller of UR5 robotic arm generates feedback messages with negligible jitter.
 In case of a 5G connection, the average IAT remains $2$ ms, because there is no packet loss.
 However, the most of the feedback messages are sticking together and form  bursts of $2$-$4$ packets typically at arrival. 
 This implies that the feedback messages queued up somewhere during its journey and arrived in bursts.
 
 \subsubsection{Command-RTT} Our remote controller generates \texttt{speedj} velocity command messages with $2$~ms period and tags each with a sequence number. In the local controller, the latest received velocity command is reflected in the $qd_{target}$ field of the feedback message. Then, the command-RTT is the time difference between command sent time and the arrival time of the first corresponding feedback message. The average command-RTT for cable was about $6$~ms. This implies that the internal cycles of the UR5 robot (i) for processing incoming commands and (ii) for generating feedback messages are asynchronous and contribute $2$~x~$1$~ms to average delay. 
 In addition, the measurement suggests that there is an addition internal delay as well, which is about $4$~ms. 
 The 5G contributed with ${\sim}18$~ms to the average RTT.

In order to implement the measurement of the command-RTT, we encoded a sequence number into the joint speed argument of the \texttt{speedj} command, which is represented using IEEE double-precision floating-point number format. From the $53$ significant bits of the last joint speed value, we reuse the $12$ less significant bits for the sequence number. This solution reduces the precision from about $16$ digits to $12$, but it is still beyond the precision of the controlled servo motors.
This mechanism also enables the system to signal additional information from remote to local controller and back, for instance, the type of the actual control command.

Based on these measurements, the remote controller can (i) monitor the quality of its closed-loop control connectivity, and (ii) even have the opportunity to adjust the settings of the remote/local control algorithm.

\subsection{Evaluation methodology}
\label{appendix:eval_methodology}

We used three focused evaluation environments. 

\subsubsection{Testbed depicted in Fig. \ref{fig:architecture}}
 
A UR5 robotic arm, a 5G network and a remote controller running in a container in the edge cloud. We used this testbed 
(i) to characterize the detailed operation of the investigated cyber-physical system and (ii) to validate the operation of the adaptive control type method. 
 
\subsubsection{Simulator}
 
For algorithm development, we used the simulated version of the UR5 robotic arm directly connected to the remote controller. The gaps were artificially added. 
 
\subsubsection{Focused simulator}
 
To evaluate the performance of the AI-based extrapolation and trajectory time-scaling method, we used an environment where only the relevant components are executed. Namely, the path planning, the trajectory generation, the trajectory time-scaling, the artificial gap generation, the AI-based extrapolation
and the calculation of the performance measures. This setup allows us to evaluate trajectories significantly faster than real-time. 

\subsection{Processing of \texttt{speedl} and \texttt{speedj} commands} \label{apendix:speedl}

The \texttt{speedl} command received by the local controller
(see Figure \ref{fig:components}) specifies the velocity in Cartesian space. 
The set point of the internal controller is internally updated with $2$~ms period. At each update, a new joint speed set point ($\underline{\dot{q}}$) 
is calculated for the internal controller. 
To determine this set point, the Cartesian space speed values ($\underline{v}$) specified by the last \texttt{speedl} command is transformed into joint space speed values $\underline{\dot{q}}$ by Jacobian matrix.
Formally, $\underline{\dot{q}} = \mathbf{J}\hspace{-1mm}\left( \underline{q} \right) \cdot \underline{v}$. 
The Jacobian matrix itself is also updated at each tick, because it depends on the actual joint position values ($\underline{q}$). 
The internal (closed-loop) controller attempts to maintain the actual joint speed set point. 

The \texttt{speedj} command updates directly the set point of the internal controller and no need for regular internal update between commands. 

\end{document}